\documentclass[manuscript]{acmart}

\usepackage{todonotes}
\usepackage{tabularx}
\usepackage{booktabs}
\usepackage{makecell}
\usepackage{multirow}
\usepackage{arydshln}
\usepackage{amsmath}
\usepackage{xspace}
\usepackage{hyperref}
\usepackage{csquotes} 
\usepackage{dirtytalk} 
\usepackage[labelformat=simple]{subcaption} 

\usepackage{cleveref} 
\crefformat{footnote}{#2\footnotemark[#1]#3}
\usepackage{tablefootnote} 

\newcommand{\eg}{\textit{e.g.}\@\xspace}
\newcommand{\ie}{\textit{i.e.}\@\xspace}
\newcommand{\etal}{\textit{et al.}\@\xspace}

\usepackage{xcolor}

\AtBeginDocument{%
  \providecommand\BibTeX{{%
    \normalfont B\kern-0.5em{\scshape i\kern-0.25em b}\kern-0.8em\TeX}}}

\copyrightyear{2021} 
\acmYear{2021} 
\setcopyright{acmlicensed}\acmConference[ASSETS '21]{The 23rd International ACM SIGACCESS Conference on Computers and Accessibility}{October 18--22, 2021}{Virtual Event, USA}
\acmBooktitle{The 23rd International ACM SIGACCESS Conference on Computers and Accessibility (ASSETS '21), October 18--22, 2021, Virtual Event, USA}
\acmPrice{15.00}
\acmDOI{10.1145/3441852.3471232}
\acmISBN{978-1-4503-8306-6/21/10}

\acmSubmissionID{9752}

\begin{document}


\title[Accessing Passersby Proxemic Signals through a Head-Worn Camera]{Accessing Passersby Proxemic Signals through a Head-Worn Camera: Opportunities and Limitations for the Blind} 

\author{Kyungjun Lee}
\email{kjlee@cs.umd.edu}
\orcid{0000-0001-8556-9113}
\affiliation{%
  \institution{University of Maryland, College Park}
  \city{College Park}
  \state{Maryland}
  \postcode{20742}
  \country{USA}
}

\author{Daisuke Sato}
\email{daisukes@cmu.edu}
\affiliation{%
  \institution{Carnegie Mellon University}
  \streetaddress{5000 Forbes Ave} 
  \city{Pittsburgh}
  \state{Pennsylvania}
  \postcode{15213}
  \country{USA}
}

\author{Saki Asakawa}
\email{saki.asakawa@nyu.edu}
\affiliation{%
  \institution{New York University}
  \streetaddress{6 MetroTech Center} 
  \city{Brooklyn}
  \state{New York}
  \postcode{11201}
  \country{USA}
}

\author{Chieko Asakawa}
\email{chiekoa@us.ibm.com}
\affiliation{%
  \institution{IBM Research}
  \streetaddress{1101 Kitchanwan Rd} 
  \city{Yorktown Heights}
  \state{New York}
  \postcode{10598}
  \country{USA}
}
\affiliation{%
  \institution{Carnegie Mellon University}
  \streetaddress{5000 Forbes Ave} 
  \city{Pittsburgh}
  \state{Pennsylvania}
  \postcode{15213}
  \country{USA}
}

\author{Hernisa Kacorri}
\email{hernisa@umd.edu}
\affiliation{%
  \institution{University of Maryland, College Park}
  \city{College Park}
  \state{Maryland}
  \postcode{20742}
  \country{USA}
}

\renewcommand{\shortauthors}{Lee, et al.}

\begin{abstract}
The spatial behavior of passersby can be critical to blind individuals to initiate interactions, preserve personal space, or practice social distancing during a pandemic. Among other use cases, wearable cameras employing computer vision can be used to extract proxemic signals of others and thus increase access to the spatial behavior of passersby for blind people. Analyzing data collected in a study with blind (N=10) and sighted (N=40) participants, we explore: (i) visual information on approaching passersby captured by a head-worn camera; (ii) pedestrian detection algorithms for extracting proxemic signals such as passerby presence, relative position, distance, and head pose; and (iii) opportunities and limitations of using wearable cameras for helping blind people access proxemics related to nearby people. Our observations and findings provide insights into dyadic behaviors for assistive pedestrian detection and lead to implications for the design of future head-worn cameras and interactions.
\end{abstract}



\begin{teaserfigure}
    \centering
    \includegraphics[width=\textwidth]{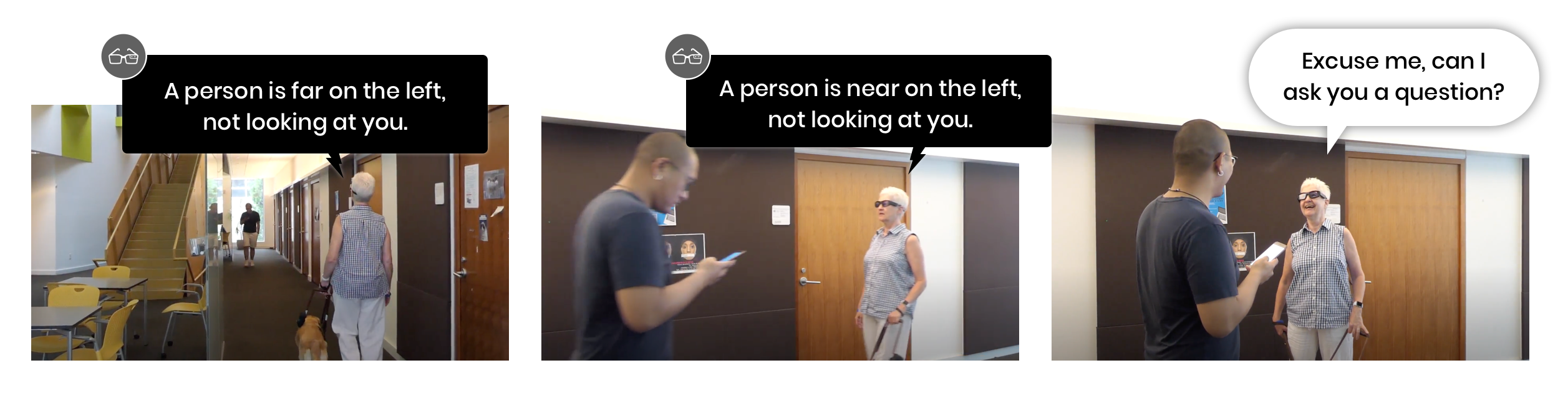}
    \caption{A pedestrian detection scenario demonstrating the data collection in our study. A blind person, wearing smart glasses with our working prototype, and a sighted person walk toward each other in a corridor. The smart glasses detect the sighted passerby, estimate his proximity, and share his relative location and head pose.}
    \label{fig:teaser}
    \Description[This figure contains three images describing how a blind user would be using smart glasses for pedestrian detection.]{From left to right, the first image shows a blind user wearing smart glasses is walking with her guide dog in a corridor.  In the same image, a pedestrian is walking toward her, but far from the blind person, which is detected by the smart glasses of the blind person.  The second image shows that the smart glasses detect the passerby near and on the left of the blind user, as their distance becomes closer.  In the last image, the blind person is then interacting with the passerby, by saying "Excuse me, can I ask you a question?"}
\end{teaserfigure}

\maketitle

\section{Introduction}
Access to the spatial behavior of passersby can be critical to blind individuals. Passersby proxemic signals, such as change in distance, stance, hip and shoulder orientation, head pose, and eye gaze, can indicate their interest in ``initiating, accepting, maintaining, terminating, or altogether avoiding social interactions''~\cite{kendon1990conducting, mead2011spatial}. More so, awareness of others' spatial behavior is essential for preserving one’s personal space. Hayduk and Mainprize~\cite{hayduk1980personal} demonstrate that the personal space of blind individuals does not differ from that of sighted individuals in size, shape, or permeability. However, their personal space is often violated as sighted passersby, perhaps in an attempt to help, touch them, or grab their mobility aids without consent~\cite{williams2014just, branham2017issomeone}. Last, health guidelines as with the recent COVID-19 pandemic~\cite{whocovid19} require one to practice social distancing by maintaining a distance from others of at least 3 feet (1 meter)\footnote{Minimum distance suggested by the World Health Organization (WHO)~\cite{whosocial}} or 6 feet (2 meters)\footnote{Minimum distance suggested by the Center for Disease Control and Prevention (CDC) in the United States~\cite{cdcsocial}}, presenting unique challenges and risks for the blind community~\cite{lighthouseguild, afb2020flatten}; mainly, due to the fact that spatial behavior of passersby and signage or markers designed to help maintain social distancing are predominantly accessible through sight and thus are, in most cases, inaccessible to blind individuals.  Other senses such as hearing and smell could be utilized perhaps to estimate passersby distance and orientation but require close proximity or quiet spaces; in noisy environments, blind people's perception of surroundings out of the range of touch is limited~\cite{feierabend2019auditory}. More so, mobility aids may work adversely in some cases (\eg, guide dogs are not trained to maintain social distancing~\cite{rickly2020covid}). 

\textbf{Why computer vision and wearable cameras.} Assistive technologies for the blind that employ wearable cameras and leverage advances in computer vision, such as pedestrian detection~\cite{stearns2018automated, lee2020pedestrian, ahmed2020investigating, grayson2020adynamic}, could provide access to spatial behavior of passersby to help blind users increase their autonomy in practicing social distancing or initiating social interaction; the latter motivated our work since it was done right before COVID-19. `Speak up' and `Embrace technology' are two of the tips that the LightHouse Guild provided to people who are blind or have low vision for safely practicing social distancing during COVID-19~\cite{lighthouseguild}, mentioning technologies such as Aira~\cite{Aira} and BeMyEyes~\cite{BeMyEyes}. However, these technologies rely on sighted people for visual assistance.  Thus, beyond challenges around cost and crowd availability, they can pose privacy risks~\cite{ahmed2015privacy, akter2020uncomfortable, stangl2020visual}.
Prior work, considering privacy concerns for parties that may get recorded, has shown that bystanders tend to be amicable toward assistive uses of wearable cameras~\cite{profita2016effect, ahmed2018up}, especially when data are not sent to servers or stored somewhere~\cite{lee2020pedestrian}.
However, little work explores how blind people capture their scenes with wearable cameras and how their data could work with computer vision models. As a result, there are few design guidelines for assistive wearable cameras for blind people.

To better understand the opportunities and challenges of employing such technologies for accessing passersby proxemic signals, we collect and analyze video frames, log data, and open-ended responses from an in-person study with blind (N=10) and sighted (N=40) participants\footnote{The data were collected right before COVID-19 with a use case scenario of blind individuals initiating social interactions with sighted people in public spaces. Out team, co-led by a blind researcher, was annotating the data when the pandemic reached our shores. Thus, we felt it was imperative to expand our annotation and analysis to include distance thresholds recommended for social distancing. After all, both scenarios share common characteristics such as need for proximity estimation and initiating interactions to maintain distancing, \ie recommended `Speak up' strategy by the LightHouse Guild~\cite{lighthouseguild}.}. As shown in Fig.~\ref{fig:teaser}, blind participants ask passersby for information while wearing smart glasses with our testbed prototype for real-time pedestrian detection.  We explore what visual information about the passerby is captured with the head-worn camera; how well pedestrian detection algorithms can extract proxemic signals in camera streams from blind people; how well the real-time estimates can support blind users at initiating interactions; and what limitations and opportunities head-worn cameras can have in terms of accessing proxemic signals of passersby.

Our exploratory study shows that there are still many limitations to overcome for such assistive technology to be effective. When the pedestrians faces are included, the results are promising. Yet, idiosyncratic movements of blind participants and the limited field of view in current wearable cameras call for new approaches in estimating passersby proximity. Moreover, analyzing images from smart glasses worn by blind participants, we observe variations in their scanning behaviors (\ie, head movement) leading to capturing different body parts of passersby, and sometimes excluding essential visual information (\ie, face), especially when they are near passersby.
Blind participants' qualitative feedback indicates that they found it easy to access proxemics of passersby via smart glasses as this form factor did not require camera aiming manipulation. They generally appreciated the potential and importance of such an wearable camera system, but commented on how experience degraded with errors. They mentioned trying to adapt to these errors by aggregating multiple estimates, instead of relying on a single one. At the same time, they shared their concern that there was no guaranteed way for them to check errors in some visual estimates.
These findings help us discuss the implications, opportunities, and challenges of an assistive wearable camera for blind people.

\section{Related Work}
Our work is informed by prior work in assistive technology for non-visual access of passersby spatial behaviors and existing approaches for estimating proximity beyond the domain of accessibility. For more context, we also share prior observations related to blind individuals' interactions with wearable and handheld cameras.

\subsection{Accessing the Spatial Behavior of Others With a Camera}
Prior work on spatial awareness for blind people with wearable cameras and computer vision has mainly focused on tasks related to navigation~\cite{fiannaca2014headlock, li2016isana, muehlbradt2017goby, noceti2019designing, thevin2020x} and object or obstacle detection~\cite{liu2015isee, li2016isana, mekhalfi2016recovering, agarwal2017lowcost, noceti2019designing}.  We find only a few attempts exploring the spatial behavior of others whom a blind individual may pass by~\cite{mcdaniel2008using, kayukawa2019bbeep} or interact with~\cite{krishna2005wearable, stearns2018automated, sarfraz2017multimodal}. For example, McDaniel \etal~\cite{mcdaniel2008using} proposed a haptic belt to notify blind users of the location of a nearby person, whose face is detected from the wearable camera. While closely related to our study, blind participants were not included in the evaluation of this system. Thus, there are no data on how well this technology would work for the intended users. Kayukawa \etal~\cite{kayukawa2019bbeep}, on the other hand, evaluate their pedestrian detection system with blind participants. However, the form factor of their system, a suitcase equipped with a camera and other sensors, is different. Thus, it is difficult to extrapolate their findings for other wearable cameras such as smart glasses that may be more sensitive to blind people's head and body movements. Designing wearable cameras for helping blind people in social interactions is a long-standing topic~\cite{krishna2005wearable, gade2009person, panchanathan2016social, morrison2021social}. Focusing on this topic, our work is also inspired by Stearns and Thieme~\cite{stearns2018automated}'s preliminary insights on the effects of camera positioning, field-of-view, and distortion for detecting people in a dynamic scene. However, they focused on interactions during a meeting (similar to~\cite{sarfraz2017multimodal}) and analyzed video frames captured by a single blindfolded user, which is not a proxy for blind people~\cite{roeder1999improved, postma2007differences, afonso2005study, tinti2006visual}.  In contrast to our work, the context of a meeting does not allow for extensive movements and distance changes --- typical tasks when proxemics of passersby are needed.
In our paper, we explore how blind people's movements and distances can lead to visual characteristics of their camera frames.

\subsection{Estimating Proximity to Others Beyond Accessibility}
Distance between people is key to understanding spatial behavior. Thus, epidemiologists and behavioral scientists have long been interested in automating the estimation of people's proximity~\cite{albrecht1982social, madan2010social}. Often their approaches employ mobile sensors~\cite{madan2010social} or Wi-Fi signals~\cite{osmani2014analysis}. For example,  Madan \etal~\cite{madan2010social} recruited participants in a residence hall and tracked foot traffic based on their phones' signals to understand how their behavior change relates to reported symptoms. While these approaches work well for the intended objective, they are limited in the context of assistive technology for real-time estimates of approaching passersby in both indoor and outdoor spaces.

Since the recent pandemic outbreak~\cite{whocovid19}, spatial proximity estimation (\ie, looking at social distancing) has gained much attention. The computer vision community started looking into this using surveillance cameras~\cite{ghodgaonkar2020analyzing, rezaei2020deepsocial} or camera-enabled autonomous robots~\cite{umd2020scifi}. Using cameras capturing people from the perspective of a third person, they monitor the crowd and detect people who violate social distancing guidelines. However, for blind users, an egocentric perspective is needed.  
Other approaches include using ultrasonic range sensors that can measure a physical distance between people wearing the sensors (\eg~\cite{david2020ultrasonic}). This line of work is promising in that the sensors consume low power, but the sensors can help detect only presence and approximate distance. They cannot provide access to other visual proxemics that blind people may care about~\cite{lee2020pedestrian}. More so, they assume that sensors are present on every person. Thus, in our work, we prioritize RGB cameras that can be worn only by the user to estimate presence, distance, and other visual proxemics.

Specifically, we borrow from pedestrian detection approaches in computer vision, but some do not really work for our context. For example, they estimate distances by looking at visual differences of the pedestrian between two images from a stereo-vision camera~\cite{kwon2005person, mustafah2012stereo}. They first track the pedestrian and then change the camera angle to position the person in the center of frames~\cite{kwon2005person}. Alternatively, they use two stationary cameras~\cite{mustafah2012stereo}. Neither case is appropriate for our objective. However, approaches that estimate the distance based on a person's face or eyes captured from an egocentric user perspective could work~\cite{flores2013camera, winarno2014development, rahman2009person, zhang2016joint}. For our testbed, we use the face detection model of Zhang \etal~\cite{zhang2016joint} to estimate a passerby's presence, distance, relative position, and head pose.

\subsection{Choosing Handheld versus Head-worn Cameras in Assistive Technologies for the Blind}
Smartphones and thus handheld cameras are the status quo for assistive technologies for the blind~\cite{bigham2010vizwiz, jayant2011supporting, vazquez2012helping, kacorri2017people, balata2015blindcamera, zhao2018face, lee2019hands, ahmetovic2020recog, troncoso2020aiguide}, though, there are still many challenges related to camera aiming~\cite{kacorri2017people, lee2019hands, bigham2010vizwiz, jayant2011supporting, vazquez2012helping, zhao2018face, lee2019revisiting, ahmetovic2020recog}, which can be trickier for a pedestrian detection scenario. 
Nonetheless, what might start as research ideas quickly translate to real-world applications~\cite{OrCam, Aira, SeeingAI, Envision, GoogleLookout, eSight, BeMyEyes, TapTapSee, LookTel, KNFB, DigitEyes}. However, the use of cameras along with computationally intensive tasks can rapidly heat up the phone and drain its battery~\cite{zhang2020smartphone, mcintosh2021what}.
Although this issue is not unique to this form factor, it can harm phone availability that blind users may rely on; many have reported experiencing anxiety when hearing battery warnings in their phones, especially at their work or during commuting~\cite{xu2019darkreader}.

With the promise of hands-free and thus more natural interactions, head-worn cameras are falling on and off the map of assistive technologies for people who are blind or have low vision~\cite{luo2006use, peli2009image, tanuwidjaja2014chroma, hwang2014augmented, zhao2015foresee, zhao2017understanding, stearns2018design, grayson2020dynamic}. This can be partially explained by hardware constraints, such as battery life and weight, limiting their commercial availability~\cite{seneviratne2017survey} which can also relate to factors such as cost, social acceptability, privacy, and form factor~\cite{profita2016effect, lee2020pedestrian}. Nonetheless, we see a few attempts in commercial assistive products~\cite{OrCam, eSight, Aira, NuEyes}, and people with visual impairments remain amicable towards this technology~\cite{zolyomi2017technology}, especially when it resembles a pair of glasses~\cite{hoogsteen2020functionality} similar to the form factor used in our testbed. However, to our knowledge, there is no prior work looking at head-worn camera interaction data from blind people, especially in the context of accessing proxemics to others. 
Although wearable cameras do not require wearers to aim a camera, including visual information of interest in the frame can still be challenging for blind people. Prior work with sighted~\cite{wolf2015effects} and blind-folded~\cite{stearns2018automated} participants indicate that the camera position can lead to different frame characteristics, though such data are not proxy for blind people's interactions nor related to the pedestrian detection scenario.
Thus, collecting and analyzing data from blind people, our work takes a step towards understanding their interactions with a head-worn camera in the context of pedestrian detection.

\section{GlAccess: A Testbed for Accessing Passersby Proxemics}
\begin{figure}
    \centering
    \includegraphics[width=0.8\columnwidth]{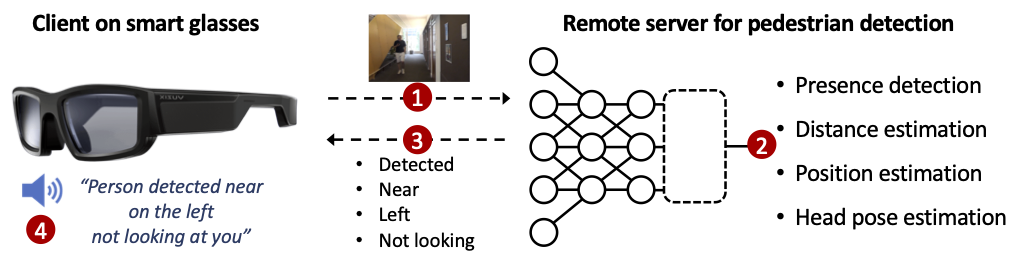}
    \caption{System diagram describing the procedure of pedestrian detection: (1) The client app on smart glasses captures an image and sends it to the remote server; (2) Once the remote server detects a passerby, it estimates the distance, position, and head pose of the passerby; (3) The output is then sent back to the smart glasses; (4) The smart glasses deliver the output to the blind user via audio.}
    \label{fig:system_diagram}
    \Description[This system diagram illustrates how the client on smart glasses communicates the remote server for pedestrian detection.]{This diagram shows that the smart glasses capture a person who appears on the left of a camera frame. This camera frame is sent to the remote server. The remote server detects the person and then estimates the distance, position, and head pose of the detected person. The detection outputs, which are Detected, Near, Left, and Not looking, are sent back to the smart glasses. The smart glasses speak out the outputs, "Person detected near on the left not looking at you."}
\end{figure}

We build a testbed called \textit{GlAccess} to explore what visual information about passersby is captured by blind individuals with a head-worn camera and how well estimates of passersby's proxemics can support initiating interactions. As shown in Fig.~\ref{fig:system_diagram}, GlAccess consists of Vuzix Blade smart glasses~\cite{VuzixBlade}, a Bluetooth earphone, and a computer vision system running on a server. With a sampling rate of one image per second, a photo is sent to the server to detect the presence of a passerby and extract proxemic signals, which are then communicated to the user through text-to-speech (TTS). To mitigate cognitive overload, estimates that remain the same as the last frame are not communicated. Vuzix Blade smart glasses have an 80-degree diagonal angle of camera view and run on Android OS. Our server has eight NVIDIA 2080 TI GPUs running on Ubuntu 16.04. We build a client app on the glasses that is connected to the server through RESTful APIs implemented with Flask~\cite{flask} and uses the IBM Watson TTS to convert the proxemic signals to audio (\eg, \textit{``a person is near on the left, not looking at you.''} as shown in Fig.~\ref{fig:teaser}). More specifically, our computer vision system estimates the following related to proxemics: 

\begin{itemize}
    \item \textbf{Presence}: A passerby presence is estimated through face detection. We employ a Multi-task Cascaded Convolutional Networks (MTCNNs)~\cite{zhang2016joint} pre-trained on 393,703 faces from the WIDER FACE dataset~\cite{yang2016wider}, which is a face detection benchmark with a high degree of variability in scale, pose, occlusion, facial expression, makeup, and illumination. The model provides a bounding box where the face is detected in the frame.
    
    \item \textbf{Distance}: For distance estimation, we adopt a linear approach that converts the height of the bounding box from the MTCNNs face detection model to the distance. Using the Vuzix Blade smart glasses, we collect a total of 240 images with six data contributors from our lab. Each contributor walks down a corridor that is approximately 66 feet (20 meters); this is the same corridor where the study takes place. We take two photos every one meter for a total of 40 photos. We train a linear interpolation function with the height of faces extracted from these photos and the ground truth distances between the contributors and the photo-taker with the smart glasses.
    This function estimates the distance as a numeric value, which is communicated either more verbosely (\eg, \textit{`16 feet away'}) or less so by stating only \textit{`near'} or \textit{`far'} based on a 16 feet (5 meters) threshold\footnote{The 6 feet (2 meters) threshold recommended from CDC for social distancing was not communicated to blind participants in the study as it preceded the COVID-19 pandemic. However, this threshold was added later in the data annotation phase and analysis as it may prove informative for future researchers working towards accessible social distancing systems.}. Participants in our study are exposed to both (\ie, experiencing each for half of their interactions). 
    
    \item \textbf{Position}: We estimate the relative position of a passerby based on the center location of the bounding box of their face detected by the MTCNNs model. A camera frame is divided in three equal-width regions, which are accordingly communicated to the user as \textit{`left'}, \textit{`middle'}, and \textit{`right'}. 
    
    \item \textbf{Head Pose}: We use head pose detection to estimate whether a passerby is looking at the user. More specifically, we build a binary classification layer on top of the MTNNs face detection model and fine-tune it with the 240 images from our internal dataset collected by the six contributors and 432 images from a publicly available head pose dataset~\cite{gourier2004estimating}. The model estimates whether a passerby is looking at the user or not.
\end{itemize}

We also simulate future teachable technologies~\cite{patel1998teachable, kacorri2017teachable} that can be personalized by the user providing faces of family and friends. In those cases, the name of the person can be communicated to the user.
Sighted members in our research group volunteered to serve this role in our user study.
\begin{itemize}
    \item\textbf{Name}: We pre-train GlAccess with a total of 144 labeled photos from five sighted lab members so that the system can recognize them in the study setting to simulate a personalized experience. This face recognition is implemented with a SVM classifier that takes face embedding data from the MTCNNs model.
\end{itemize}

\section{User Study and Data Collection}

\begin{figure*}[t]
    \centering
    \includegraphics[width=0.96\textwidth]{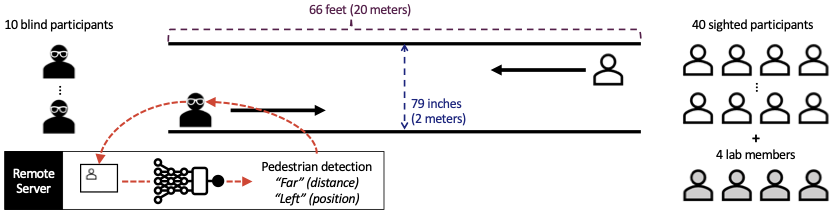}
    \caption{Diagram of study procedure.  A blind participant walked eight times in a corridor that is 66 feet (20 meters) long and 79 inches (2 meters) wide. Each time a different sighted person walked towards them. In four of the eight times, those individuals were sighted study participants. In the other four, they were sighted members in our research lab. In addition to the proxemic signals, GlAccess can recognize the lab members, simulating future teachable applications~\cite{kacorri2017teachable} where users can personalize the model for familiar faces.}
    \label{fig:study_diagram}
    \Description[This figure illustrates how the user study was done in a hallway in which a blind participant and a sighted participant were walking toward each other.]{The corridor has 66 feet (20 meters) of length and 79 inches (2 meters) of width.  At each end of this corridor, a blind participant and a sighted participant start walking toward each other.  The blind participant is wearing smart glasses that send a camera frame to a server that detects a pedestrian and extracts the distance and position of the person.  This detection output is then delivered back to the blind user through the smart glasses. 10 blind participants were recruited, and 4 sighted participants were recruited for each blind participant.}
\end{figure*}

To explore the potential and limitations of head-worn cameras for accessing passersby proxemics, we conducted an in-person study with blind and sighted participants (under IRB \#$STUDY2019\_00000294$). We employed a scenario, illustrated in Fig.~\ref{fig:study_diagram}, where blind participants wearing GlAccess were asked to initiate interactions with sighted passersby in an indoor public space. Given the task of asking someone for the number of a nearby office, they walked down a corridor where sighted participants were coming in the opposite direction.

\subsection{Participants}
We recruited ten blind participants; nine were totally blind, and one was legally blind. On average, blind participants were 63.6 years old ($SD$ = 7.6). Five self-identified as female and five as male.  Three participants (P4, P8, P10) mentioned having light perception, and four participants reported having experience with wearable cameras such as Aira~\cite{Aira} with frequencies shown in Table~\ref{tab:participants}.
Responses to technology attitude questions\footnote{\label{note:questions}Questionnaire based on~\cite{rosen2013media} is available at \url{https://iamlabumd.github.io/assets2021_lee}: -3 indicates the most negative, 0 neutral, and 3 the most positive.} indicate that they are positive about technology in general ($mean$ = 2.02, $SD$ = 0.42) and wearable technology such as smartglasses ($mean$ = 1.89, $SD$ = 0.90).

We also recruited 40 sighted participants to serve as passersby in the study --- four sighted participants, one at a time, passed by a blind participant. On average, sighted participants were 25.3 years old ($SD$ = 4.8).  Twenty-four self-identified as male and fifteen as female, and one chose not to disclose their gender. Participants' responses to technology attitude questions\cref{note:questions} indicate that they were slightly positive about wearable technology ($mean$ = 1.08, $SD$ = 0.95).

\begin{table}[t]
    \small
    \centering
    \caption{Demographic information of our participants.  Participants with light perception are marked with an asterisk in \textit{Vision level}.}
    \begin{tabular}{@{}cccccc@{}}
    \toprule
    PID & Gender & Age & Vision level & Onset age & Experience with wearable cameras \\
    \midrule
    P1 & Female & 71 & Totally blind & 8 & Aira (1 year 6 months; a few times a month) \\
    P2 & Male & 66 & Totally blind & Birth & Audio-glasses (4 months; twice a month) \\
    P3 & Male & 59 & Totally blind & Birth & None \\
    P4 & Female & 66 & Totally blind* & 49 & None \\
    P5 & Female & 65 & Totally blind & Birth & None \\
    P6 & Female & 73 & Totally blind & Birth & None \\
    P7 & Female & 49 & Totally blind & 14 & None \\
    P8 & Male & 60 & Legally blind* & 56 & Aira (1 month; once a month) \\
    P9 & Male & 56 & Totally blind & 26 & None \\
    P10 & Male & 71 & Totally blind* & 41 & Aira (3 months; a few times a week) \\
    \bottomrule
    \end{tabular}
    \label{tab:participants}
\end{table}

\subsection{Procedure}
We shared the consent form with blind participants via email to provide ample time to read it, ask questions, and consent prior to the study. Upon arrival, the experimenter asked the participants to complete short questionnaires on demographics, prior experience with wearable cameras, and attitude towards technology. The experimenter read the questions; participants answered them verbally. 
Blind participants first heard about the system and then practiced walking with it in the study environment and detecting the experimenter as he approached in the opposite direction.  During the main data collection session, blind participants were told to walk in the corridor and ask a person, if detected, about the office number nearby. On the other hand, sighted participants, one at a time, were told to simply walk in a corridor with no information about the presence of blind participants in the study.
As illustrated in Fig.~\ref{fig:study_diagram}, a blind participant walked in the corridor eight times --- four times with four sighted participants, respectively, and four times with four sighted members of our research lab, respectively. Our study was designed not to extend beyond 2 hours.

All sighted participants were recruited on site and thus strangers to blind participants; blind participants did not meet any sighted participants before the study. A stationary camera was used to record the dyadic interactions in addition to the head-worn camera and the server's logs.  Both blind and sighted participants consented to this data collection prior to the study --- sighted participants were told that there will be camera recording in the study, which may include their face, but these images will not be publicly available without being anonymized. The study ended with a semi-structured interview for blind participants with questions\cref{note:questions} created for eliciting their experience and suggestions towards the form factor, proxemics feedback delivery, error interaction, and factors impacting future uses of such head-worn cameras.

\subsection{Data Collection, Annotation, and Analysis}
We collected 183-minute-long session recordings from a stationary camera, more than 1,700 camera frames from the smart glasses testbed sent to the server, estimation logs on proxemic signals of sighted passersby that were communicated back to blind participants, and 259-minute-long audio recordings of the post-study interview.

\subsubsection{Data annotation}

\begin{table}[t]
    \small
    \centering
    \caption{Attributes in our annotation.}
    \resizebox{0.95\columnwidth}{!}{
    \begin{tabular}{c|c|c|c|c|c}
    \toprule
     & \textbf{Presence} & \textbf{Position} & \textbf{Distance} & \textbf{Head pose} & \textbf{Interaction} \\
    \hline
    Annotation & \makecell{Which body part of\\the passerby is included?}  & \makecell{Where is the\\passerby located?} & \makecell{How far is the passerby\\from the user?} & \makecell{Is the passerby\\looking at user?} & \makecell{Who is speaking?} \\
    \hline
    Value & head/torso/arms/legs & left/middle/right & <6ft./near/far & yes/no & blind/sighted \\
    \bottomrule
    \end{tabular}
    }
    \label{tab:annotation}
\end{table}

We first annotated all camera frames captured by the smart glasses worn by blind participants. As shown in Table~\ref{tab:annotation}, annotation attributes include the presence of a passerby in the video frame, their relative position and distance from the blind participant, whether their head pose indicates that they were looking towards the participant, and whether there was an interaction.
Annotations for the binary attributes, \textit{presence}, \textit{head pose}, and \textit{interaction}, were provided by one researcher as there was no ambiguity.  For \textit{presence}, the annotator marked whether the \textit{head}, \textit{torso}, \textit{arm(s)}, and \textit{leg(s)} of the sighted passerby were captured in the camera frames. Regarding \textit{head pose}, the annotator checked whether the sighted passerby was looking at the blind user. As for \textit{interaction}, the annotator marked whether the blind or the sighted participant was speaking while consulting the video recordings from the stationary camera. Annotations for \textit{distance} and \textit{position} were provided by two researchers, who first annotated independently and then resolved disagreements.  The initial annotations from the two researchers achieved a Cohen's kappa score of 0.89 and 0.90 for distance and position, respectively. \textit{Distance} was annotated as \textit{< 6ft.}: within 6 feet (2 meters), \textit{near}: farther than 6 feet (2 meters) but within 16 feet (5 meters); \textit{far}: farther than 16 feet (5 meters).
Distance in frames where the passerby was not included was annotated as \textit{n/a}.
As for \textit{position}, the annotators checked whether the person was on the \textit{left}, \textit{middle}, or \textit{right} of the frame. This relative position was determined based on the location of the center of the person's face within the camera frame. When the face was not included, the center of included body part was used.
The annotators resolved all annotation disagreements together at the end.

\subsubsection{Data analysis}
For our \textit{quantitative analysis}, we mainly used descriptive analysis on our annotated data to see what visual information about passersby blind participants captured using the smart glasses and how well our system performed to extract proxemic signals (\ie, the presence, position, distance, and head pose of passersby). We adopted the common performance metrics for machine learning models (\ie, F1 score, precision, and recall) to measure the performance of our system in terms of extracting proxemic signals.

For our \textit{qualitative analysis}, we first transcribed blind participants' responses to the open-ended interview questions. Then, we conducted a thematic analysis~\cite{braun2006using} on the open-ended interview transcripts. One researcher first went through the interview data to formulate emerging themes in a codebook. Then, two researchers iteratively refined the codebook together while applying codes to the transcripts and addressing disagreements. The final codebook had a 3-level hierarchical structure --- \ie, level-1: 11 themes; level-2: 20 themes; and level-3: 71 themes.

\subsection{Limitations}
We simplified our user study to have only one passerby in a corridor since our prototype system was able to detect only one person at a time. In the future, designers should consider handling a case where more than one passerby appears. When more than one passerby is included in a camera frame, a system needs to detect a person of the most interest to blind users and then delivering the person's proxemics to the users. Such a method should help the users perceive the output, efficiently and quickly. For example, detecting a person within a specific distance range could be a naive but straightforward way.

\section{Observations and Findings}
Although our data collection scenario is somewhat limited, the images, videos, and logs collected from 80 pairs of blind and sighted individuals allow us to account for variations within a single blind participant and across multiple blind participants. Our observations, contextualized with blind participants' feedback, provide rich insights on the potential and limitations of head-worn cameras.

\subsection{Visual Information Accessed by a Head-worn Camera}
\label{sec:quant_visual_info}
In a real-world setting, where both blind users and other nearby people are constantly in motion, it is important to understand what is being captured by assistive wearable cameras --- smart glasses in our case. What is visible dictates what kind of visual information regarding passersby's proxemics can be extracted. In this section, results are based on our analysis of 3,175 camera frames from the head-worn camera and those ground-truth annotations.

\begin{figure*}
    \centering
    \begin{subfigure}{0.9\textwidth}
        \includegraphics[width=\textwidth]{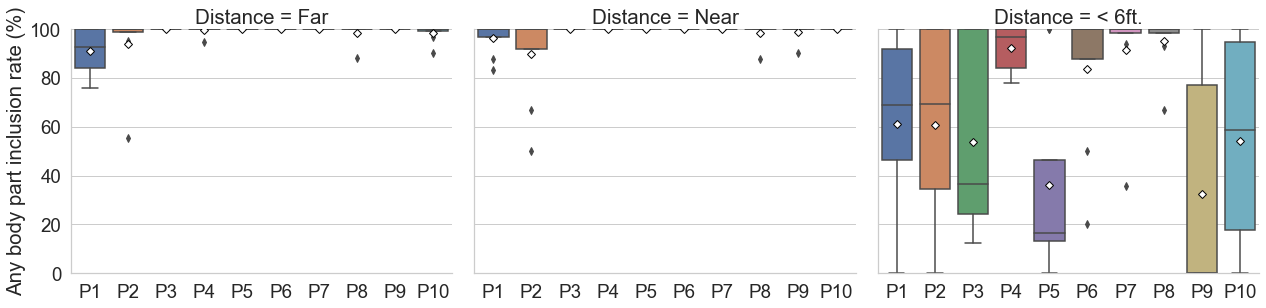}
        \caption{Inclusion of any body part of passersby in camera frames per distance category.}
        \label{fig:any_inclusion_ratio}
        \Description[This figure shows that blind participants captured any body part of pedestrians well when participants farther than 6 feet (2 meters), but many of them started excluding the pedestrians from their camera frames when within 6 feet (2 meters).]{When passersby were farther than 6 feet (2 meters) from blind participants, all the blind participants included any body part of passersby, on average, at least in 89.5\% of their camera frames. However, when the passersby were within 6 feet (2 meters), the percentage of including any body part in camera frames is, on average, 61\% (std.=40.4) for P1, 61\% (std.=42.3) for P2, 54\% (std.=39.5) for P3, 92\% (std.=9.9) for P4, 36\% (std.=40.2) for P5, 84\% (std.=31.1) for P6, 91\% (std.=22.5) for P7, 95\% (std.=11.7) for P8, 32\% (std.=45.1) for P9, 54\% (std.=42.4) for P10.}
    \end{subfigure}
    \par\bigskip
    \begin{subfigure}{0.9\textwidth}
        \includegraphics[width=\textwidth]{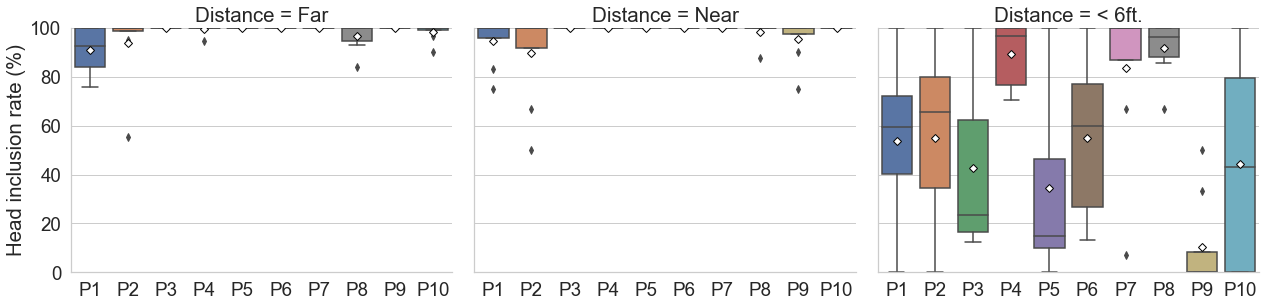}
        \caption{Inclusion of the head of passersby in camera frames per distance category.}
        \label{fig:head_inclusion_ratio}
        \Description[This figure shows that blind participants captured the head of pedestrians well when participants farther than 6 feet (2 meters), but many of them started excluding the head of pedestrians from their camera frames when within 6 feet (2 meters).]{When passersby were farther than 6 feet (2 meters) from blind participants, all the blind participants included the head of passersby, on average, in at least 89.5\% of their camera frames. However, when the passersby were within 6 feet (2 meters), the rate of including the head in camera frames varies by participants. More specifically, the inclusion ratio is, on average, 54\% (std.=36.7) for P1, 55\% (std.=37.4) for P2, 43\% (std.=37.2) for P3, 89\% (std.=13.5) for P4, 35\% (std.=41.2) for P5, 55\% (std.=32.3) for P6, 83\% (std.=32.9) for P7, 92\% (std.=11.6) for P8, 10\% (std.=19.8) for P9, 44\% (std.=42.1) for P10.}
    \end{subfigure}
    \caption{Diagrams presenting inclusion rates of any body part or head of passersby in camera frames per distance. A relatively high percentage of camera frames capture a passerby (face and body) at a distance larger than 6 feet in a corridor that is 79 inches wide; this can vary across blind participants with some outlier cases being as low as 50\%. Within 6 feet it can get unpredictable with faces often being included less.}
    \label{fig:inclusion_ratio}
    \Description[This figure consists of two subfigures.]{One on the top shows how well blind participants included any body part of pedestrians in their camera frames, while another on the bottom shows how well the participants captured the head of passersby in the camera frames.}
\end{figure*}

\subsubsection{At what distance is a passerby captured by the the camera?}

Fig.~\ref{fig:inclusion_ratio} presents that a relatively high percentage of camera frames captured a passerby at a distance larger than 6 feet in a corridor that was 79 inches wide; this varied across blind participants with some outlier cases being as low as 50\%. Within 6 feet, blind participants seemed to have different tendencies of including either the face or any body part of passersby, which average inclusion rates tend to be lower than those when passersby were more than 6 feet away.  When sighted passersby were more than 16 feet away (\textit{far}), part of their body and head were included on average in 98\% ($SD$ = 3.5) and 98\% ($SD$ = 3.6) of the camera frames, respectively. When sighted passersby were less than 16 feet but more than 6 feet away (\textit{near}), the inclusion ratios of any body part and head were 98\% ($SD$ = 3.5) and 98\% ($SD$ = 4.3), respectively.
However, when passersby were within 6 feet (2 meters), there was a quick drop in the inclusion rate with high variability across the dyads --- passersby's head and any body part were included on average in 56\% ($SD$ = 39.1) and 66\% ($SD$ = 39.6) of the camera frames, respectively. For example, the camera of P9, who tended to walk fast, consistently did not capture the passerby at this distance.

Beyond fast pace, there are other characteristics that can contribute to low inclusion rates such as veering~\cite{guth1994veering} and scanning head movements. We observed little veering, perhaps due to hallway acoustics. However, we observed that many blind participants tended to move their head side to side while walking and more so while interacting with sighted participants to ask them about the nearby office number. This behavior in combination with the low sampling rate of one frame per second seemed to contribute to exclusion of the passersby from the camera frames even when they were in front or next to the blind participants.
These findings suggest a departure from naive fixed-rate time-based mechanism (\ie 1 frame/sec) to more efficient dynamic sampling (\eg increasing frames on proximity) as well as fields of view and computer vision models that can account for such movements.

\subsubsection{What proportion of a passerby is being captured by the camera while approaching versus while interacting?} 
\begin{figure*}[t]
    \centering
    \includegraphics[width=\textwidth]{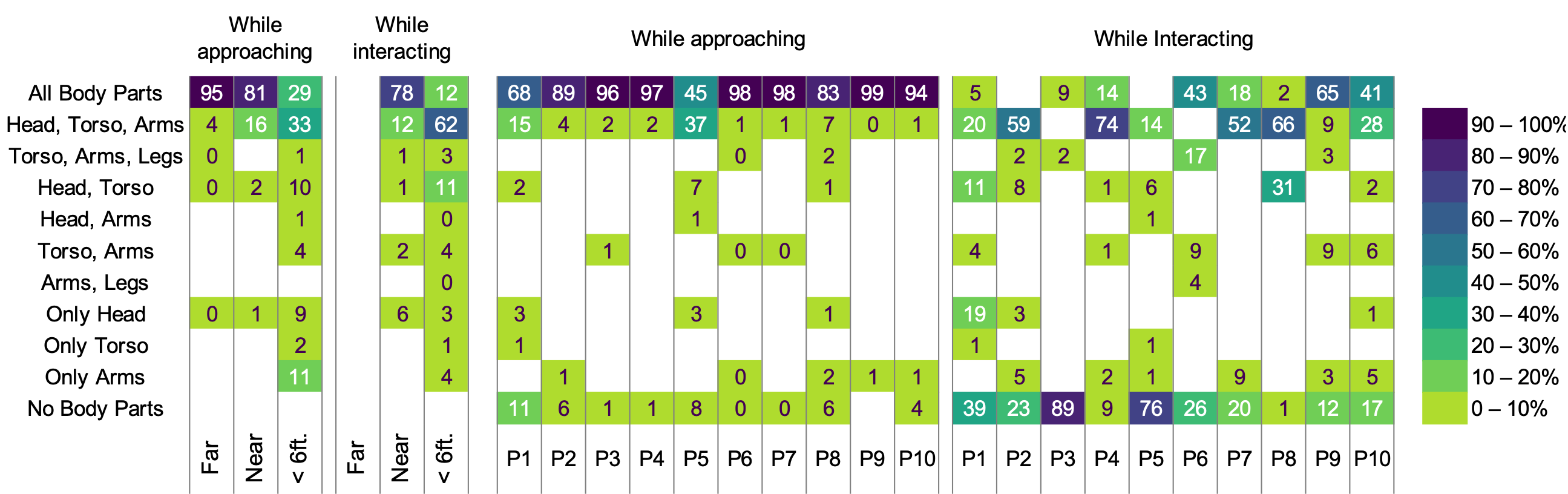}
    \caption{Heatmap diagram presenting the rates of body inclusion patterns in frames across distances, interactions, and participants.}
    \label{fig:visual_chars}
    \Description[This diagram shows the inclusion rate of each pattern per blind participant as well as per distance category in the cases of while approaching and while interacting, respectively.]{Note that the sum of the rates in each participant or each distance category is 100.  In the case of while approaching, P1 shows 67.6\% of their camera frames including all body parts, 15.1\% of including head, torso, and arms, and 21\% of including head and torso, 3.4\% of including only head, 8\% of including only torso, and 10.9\% of including no body parts.  P2 shows 88.7\% of including all body parts, 4.3\% of including head, torso, and arms, 1.4\% of including only arms, and 5.7\% of including no body parts.  P3 shows 95.7\% of including all body parts, 2.4\% of including head, torso, and arms, 1.2\% of including torso and arms, and 0.6\% of including no body parts.  P4 shows 97.1\% of including all body parts, 2.4\% of including head, torso, and arms, and 0.6\% of including no body parts.  P5 shows 45.2\% of including all body parts, 37.1\% of including head, torso, and arms, 6.7\% of including head and torso, 0.5\% of including head and arms, 2.9\% of including only head, and 7.6\% of including no body parts.  P6 shows 97.6\% of including all body parts, 0.8\% of including head, torso, and arms, 0.4\% of including torso, arms, and legs, 0.4\% of including torso and arms, 0.4\% of including only arms, and 0.4\% of including no body parts.  P7 shows 97.8\% of including all body parts, 1.3\% of including head, torso, and arms, 0.4\% of including torso and arms, and 0.4\% of including no body parts.  P8 shows 83.1\% of including all body parts, 7\% of including head, torso, and arms, 1.5\% of including torso, arms, and legs, 0.5\% of including head and torso, 0.5\% of including only head, 1.5\% of including only arms, and 6\% of including no body parts.  P9 shows 98.7\% of including all body parts, 0.4\% of including head, torso, and arms, 0.9\% of including only arms.  P10 shows 94.4\% of including all body parts, 0.5\% of including head, torso, and arms, 0.9\% of including only arms, and 4.2\% of including no body parts.  With respect to the proxemic zones, for < 6ft.,  it shows 28.6\% of including all body parts, 33\% of including head, torso, and arms, 1.1\% of including torso, arms, and legs, 9.9\% of including head and torso, 1.1\% of including head and arms, 4.4\% of including torso and arms, 8.8\% of including only head,  2.2\% of including only torso, and 11\% of including only arms; for near, it shows 81.2\% of including all body parts, 15.8\% of including head, torso, and arms, 1.9\% of including head and torso, 1.1\% of including only head; for far, 95.3\% of including all body parts, 4.1\% of including head, torso, and arms, 0.2\% of including torso, arms, and legs, 0.3\% of including head and torso, 0.2\% of including only head.
    In the case of while interacting, P1 shows 5.4\% of camera frames including all body parts, 20.3\% of including head, torso, and arms, 10.8\% of including head and torso, 4.1\% of including torso and arms, 18.9\% of including only head, 1.4\% of including only torso, and 39.2\% of including no body parts.  P2 shows 59\% of including head, torso, and arms, 1.6\% of including torso, arms, and legs, 8.2\% of including head and torso, 3.3\% of including only head, 4.9\% of including only arms, and 23\% of including no body parts.  P3 shows 9.1\% of including all body parts, 2.3\% of including torso, arms, and legs, and 88.6\% of including no body parts.  P4 shows 14.4\% of including all body parts, 73.6\% of including head, torso, and arms, 0.8\% of including head and torso, 0.8\% of including torso and arms, 1.6\% of including only arms, and 8.8\% of including no body parts.  P5 shows 13.9\% of including head, torso, and arms, 5.6\% of including head and torso, 1.4\% of including head and arms, 1.4\% of including only torso, 1.4\% of including only arms, and 76.4\% of including no body parts.  P6 shows 43.4\% of including all body parts, 17\% of including torso, arms, and legs, 9.4\% of including torso and arms, 3.8\% of including arms and legs, and 26.4\% of including no body parts.  P7 shows 18.2\% of including all body parts, 52.3\% of including head, torso, and arms, 9.1\% of including only arms, and 20.5\% of including no body parts.  P8 shows 2.3\% of including all body parts, 65.5\% of including head, torso, and arms, 31\% of including head and torso, and 1.1\% of including no body parts.  P9 shows 64.7\% of including all body parts, 8.8\% of including head, torso, and arms, 2.9\% of including torso, arms, and legs, 8.8\% of including torso and arms, 2.9\% of including only arms, and 11.8\% of including no body parts.  P10 shows 40.9\% of including all body parts, 28.4\% of including head, torso, and arms, 2.3\% of including head and torso, 5.7\% of including torso and arms, 1.1\% of including only head, 4.5\% of including only arms, and 17\% of including no body parts.  With respect to the proxemic zones, for < 6ft., it shows 12.3\% of including all body parts, 62\% of including head, torso, and arms, 2.7\% of including torso, arms, and legs, 11.4\% of including head and torso, 0.2\% of including head and arms, 3.7\% of including torso and arms, 0.5\% of including arms and legs, 3\% of including only head,  0.5\% of including only torso, and 3.7\% of including only arms; for near, it shows 77.9\% of including all body parts, 11.6\% of including head, torso, and arms, 1.2\% of including torso, arms, and legs, 1.2\% of including head and torso, 2.3\% of including torso and arms, and 5.8\% of including only head; for far, none image was captured.}
\end{figure*}

The types of visual information and proxemic signals desired by a blind user can differ by proximity to a passerby and whether that person is merely approaching or interacting with them. Nonetheless, what can be accessible will depend on what part of the passersby body is captured by the camera (\eg, head, torso, and hands). In our dataset, we identified 11 distinct patterns. For example, a \textit{(Head, Torso)} pattern indicates that only the head and torso of a passerby were visible in a camera frame. Fig.~\ref{fig:visual_chars} shows the average rate for each pattern and whether a passerby was \textit{approaching} or \textit{interacting} with a blind participant. It also includes a break-down of rates across participants. 

When passersby were more than 16 feet away (\textit{far}) and approaching, blind participants tended to capture the passersby in their entirety in more than 81\% of the overall frames. This number droped to 29\% in the distance within 6 feet, making it difficult to estimate some of their proxemics, but the head presence remained high at 82\%. These seemed to be consistent among the participants, except P1 and P5 whose rates were not only lower for including passersby in their entirety (68\% and 45\%, respectively) but also higher for not including them at all (11\% and 8\%, respectively).

We also analyzed the frames collected during blind participants' interactions with sighted passersby, typically within 16 feet (5 meters). Blind participants tended to capture the passersby in their entirety in about 78\% of the overall frames for distances between 6 and 16 feet (2--5 meters). Closer than 6 feet, the number quickly dropped to 12\%. More importantly, we observed a higher diversity in the inclusion rate across blind participants for what their camera captured when they were interacting with a passersby within 16 feet. Indeed, the breakdown of rates shows that all except P8, who was legally blind, and P4, who reported having a light perception and onset age of 49, did not include the passersby or their head in more than a quarter of their camera frames. For P3, P5, and P6, who had no prior experience with wearable cameras and reported onset age at birth, these rates were even higher reaching up to 91\%. When inspecting their videos from the static camera, we find that these participants were indeed next to the passersby interacting with them but they were not facing them.

\subsection{Errors in Estimating Passersby Presence, Distance, Position, and Head Pose}
\label{sec:quant_ret_errors}

We evaluate how well models that rely on face detection can estimate passersby' proxemic signals (\ie, \textit{presence}, \textit{position}, \textit{distance}, and \textit{head pose}) in camera frames from blind participants. Specifically, using visual estimates and those ground-truth annotations, we report results in terms of precision, recall, and F1-score (Table~\ref{tab:performance}).  These results capture blind participants' experience with the testbed, and thus provide context for interpreting their responses related to machine learning errors.

Table~\ref{tab:performance} presents that, when the presence of a passerby was detected, there was actually a passerby (precision=1.0)\footnote{Such a high precision is not surprising since this detection task in our study was an imbalanced classification problem: there were two classes --- \textit{presence} and \textit{no presence} --- with the \textit{presence} representing the overwhelming majority of the data points since there was always a passerby coming towards the blind participants. More so, there were no visual elements in the study environment that could confuse the model (\eg, photos of people on the walls).}. However, the model's ability to find all the camera frames where a passerby was present is low (recall=0.33) --- even when their head was visible in the camera frame (recall=0.34). Our visual inspection on these false negatives informed that, for the majority (80\%), this was due to the person being too far (> 16 feet); thus, their face was too small for the model to detect, raising questions such as \textit{How far away should a system detect people?}

\begin{table}[t]
    \small
    \centering
    \caption{Performance of our prototype system on the camera frames from blind participants.}
    \begin{tabular}{@{}c|c|ccc@{}}
    \toprule
    \textbf{Attribute} & \textbf{Comparison} & \textbf{Precision} & \textbf{Recall} & \textbf{F1-score} \\
    \midrule
    \multirow{2}{*}{Presence} & across all frames where a body part was present  & 1.0 & 0.33 & 0.49 \\
                              & across all frames where head was present & 1.0 & 0.34 & 0.51 \\
    \midrule
    \multirow{2}{*}{Position} & across all frames where someone was present & 0.90 & 0.30 & 0.45 \\
                              & across all frames where someone was detected & 0.90 & 0.84 & 0.87 \\
    \midrule
    \multirow{2}{*}{Distance} & across all frames where someone was present & 0.62 & 0.25 & 0.35 \\
                              & across all frames where someone was detected & 0.62 & 0.66 & 0.58 \\
    \midrule
    Head pose & across all frames where someone was detected & 0.77 & 0.77 & 0.77 \\
    \bottomrule
    \end{tabular}
    \label{tab:performance}
\end{table}

Since the estimation of \textit{position}, \textit{distance}, \textit{head pose} depended on the \textit{presence} detection, we reported metrics on both (i) frames where passersby were present and (ii) frames where they were first detected. Once passersby were detected, the estimation of their relative position worked relatively well (precision=0.90; recall=0.84).  However, the head pose estimation (\ie, whether the passerby is looking at the user) was not as accurate (precision=0.77; recall=0.77). The distance estimation was also challenging (precision=0.62; recall=0.66). This might be partially due to the limited training data, both in size and in diversity. A passerby's head size in an image, a proxy for estimating the distance, can depend on the blind user's height and camera angle.

\subsection{Participant Feedback on Accessing Proxemics}
\label{sec:qual_accessing_proxemics}

The majority of blind participants (seven) appreciated having access to proxemics about nearby people through the smart glasses. P4 highlighted, \textit{``I liked having the [proxemic] information. That means a lot when I know that there's a person, or how close they are to you or where they are[.] It's a good thing to be able to know who's there and that there's somebody there. That's important.''}
Blind participants also shared certain situations where they can benefit from such assistive cameras (\eg, when nearby people do not provide enough non-visual cues). P2 highlighted, \textit{``People are sometimes quite, and [I] don't know if they are approaching.''}
Some participants remarked that having access to proxemics related to a near passerby can help them make their decision proactively. For example, P7 mentioned, \textit{``With the distance estimations, it gives [me] a couple of cues within range, where [I] can make a decision `Do I want to do something or not?' ''}

Participants also envisioned how these visual estimations could enhance their social engagement. For example, P1 said, \textit{``It was novel [and] interesting to have it tell me things that I normally wouldn't even be paying attention to[.] I think it would be a great social interaction thing,''} and P4, \textit{``It means a lot to be able to know who it is [and] who's there --- that's a big deal to me for social interaction.''}
Some participants see its potential in helping them predict the outcome of their attempted social interaction by saying, \textit{``If you know somebody is in front of you and looking [at you] or you know [where] to look (left or right), you can probably get their attention more easily''} (P5), and \textit{``[GlAccess] certainly made it easy to know, when there was a potential [social interaction] --- you didn't talk to the air[.] I would say that I like the ability to possibly predict the outcome of an attempted interaction''} (P7).
They also envisioned using proxemics detection and person recognition to initiate their communication. P6 said, \textit{``Walk down the hall and be able to greet a person by name when they hadn't spoken to me; that would be cool,''} and P7, \textit{``Because [I] know that there is at least a person and how far away they are, [I'm] not simply chattering with air, which just occurred to me a couple times [in my daily life].''}


On the other hand, our analysis also reveals potential reasons why blind participants might not pay much attention to visual estimates. Since such visual information was mostly inaccessible in their daily lives, they did not rely on the visual information and also was unsure how to utilize them. More specifically, P7 and P9 remarked, \textit{``[Getting visual information] never occurred to me. [Not sure if] I would want or need anything more than just listening to the voices around me and hearing the walking around me because I've never had that experience,''} and \textit{``It's visual feedback that I'm not used to, I never functioned with, so I don't know, I'm not used to processing or analyzing things in a visual way. I love details, but I just never often think about them much related to the people,''} respectively.
To address this issue, participants suggested providing training that helps blind users interpret such visual estimates as P7 said, \textit{``I would need maybe to be trained on what to do [with the visual feedback].''}


\subsection{Participant Feedback on the Form Factor}
\label{sec:qual_form_factor}
Six participants commented on how easy it was to have the smart glasses on and access information, saying \textit{``Just wearing the glasses and knowing that would give you feedback''} (P2); \textit{``Just have to wear it and listen''} (P4); and \textit{``Once [I] put on and just walk''} (P8). Some supported their statements by proving a rationale such as the head being a natural way to aim towards areas of interest and the fact that they didn't have to do any manipulations.  For example, P4 and P9 commented, \textit{``If it's just straight, then [I'm] gonna see what's straight in front of [me]. That's a good benefit[.] Seeing what directly in front of [me] is a bonus''} and \textit{``I could turn my head to investigate or find out what's to my right or my left --- that part was kind of nice,''} respectively.  P7 and P10 commented on the manipulation, saying \textit{``You don't have a lot of controlling and manipulating to do''} and \textit{``And [it] doesn't take any manipulation --- it just feeds new information. And that was easy. It took no interaction manually changed,''} respectively.

However, anticipating that some control would be eventually needed, P8 tried to envision how one could access the control panel for the glasses and suggested them being paired with the phone or support voice commands by stating \textit{``Altering [it] would be an interface with your phone so that I could [control] it using the voice over on my phone, or using Siri to do it. The voice assistant [would] let you control.''} Only one participant (P1) suggested deploying similar functionalities into one's phone \textit{``And just use your rear facing camera and have it as a wearable, a pouch or something like that. Then you just flip that program on [the phone] when you want to use it.''}

Other suggestions focused on the sound. At the time, Vuzix Blade did not have a built-in speaker, and their shape conflicted with bone conduction headphones. Four participants quickly pointed out this limitation with statements such as \textit{``Bone conduction [would] seem to work [with GlAccess] really good''} (P4) and \textit{``[Need to keep] your ear canals open there for being able to hear traffic or whatever''} (P8).
P4 suggested extending the field of view for capturing more visual information, stating \textit{``I wonder if the field of vision for the camera could go a little bit wider because then it would catch somebody coming around the side more[.]''}

\subsection{Participant Feedback on Interactions with the System}
\label{sec:qual_interactions}

Four participants shared concerns on outside noises blocking out the system's audio feedback, or vice versa. P4 mentioned, \textit{``The only time it could be harder [to listen to the system feedback] is in a situation with lots of noise or something. If it's very noisy, then it's going to be hard to hear,''} and \textit{``There's a limit to how much you can process because you're walking. When we're walking[,] we're doing more than just listening to the [the feedback.] We listen to [others]. Listening is big [to us].''}
Although GlAccess reported only feedback changed between two contiguous frames, this is possibly why they wanted to control the verbosity of the system feedback. P7 said, \textit{``It was reacting to you every couple seconds or literally every second. Until I'm looking for something specific, it [is] a little overwhelming to have all of it coming in at the same time.''}

Moreover, blind participants emphasized that they would need to access such proxemic signals in real time, especially when they are walking. They mentioned being misguided by stale estimations caused by network latency at times --- P9 commented on this latency issue, \textit{``I know about the latency in it. Because when [a person] and I were talking, I knew [the person] walked away, [but GlAccess] still reported that [the person was] in front of me.''}
To minimize the effects of the latency, some reported that they made quick adaptation, such as \textit{``walking slow[ly]''} (P1) or \textit{``stop[ping] and paus[ing]''} (P10), during the study. Nonetheless, the latency issue would make the use of such assistive wearable cameras impractical in the real world, as P8 put, \textit{``If I'm walking at a normal pace [and] it's telling me somebody is near, I'm probably going to be 10 to 15 feet past [due to the latency].''}


Participants also suggested providing options for customizing the feedback delivery. For example, P7 wanted the system to detect moments for pausing or muting the feedback delivery, saying \textit{``Once a conversation is going on, it's not reporting everyone else around; [otherwise, it's] distracting [me].''}
Also, four blind participants commented on verbosity control, which we explored a bit in our study by implementing two ways of communicating the distance estimation. P2 mentioned, \textit{``I like that you had options [for verbosity], and [I] could choose that.''} At the same time, we observe that blind people may have different definitions of spatial perception --- \ie, the definition of \textit{far} and \textit{near}. P4 remarked on this, \textit{`` `Far'--`near' thing wasn't so good. I liked hearing the the number of feet for the distance or my concept of far is different than your concept of far and near[.]''}

\subsection{Participant Feedback on Estimates and Errors}
\label{sec:qual_estimates_errors}
Seven blind participants mentioned being confident of interpreting person recognition output. It seems that they gained their trust in the person recognition feature since they were able to confirm the recognition result with a person's voice. P2 said, \textit{``I learned [a lab member]'s voice, and when it said `[lab member]', it was her; or at least I thought it was.''}
Similar to any other machine learning models, however, our prototype system also suffered from errors such as false negatives (Section~\ref{sec:quant_ret_errors}).
Participants seemed to be familiar with technical devices making errors and understood that it is part of technology improvement, as P1 commented, \textit{``If it didn't work, then I figured that's just part of the the system growing, so that's okay. It's just part of learning how to use it.''}
To prevent from making their decision based on errors, six participants mentioned aggregating multiple estimates from the system, instead of relying on a singular estimate, after realizing the possibility of errors in visual estimates. P9 highlighted, \textit{``If [the] information was not what I felt either incorrect or not certain, I would wait and give the system a chance to focus better or to process it differently.''}
Even with errors in visual estimates, blind participants said that it can help them know that \textit{``there was somebody [at least]''} (P6). On the other hand, seven participants pointed out that there would be no guaranteed way for them to verify the visual estimates --- \ie, detecting errors in visual estimates is inaccessible.
For example, P3 did not know if there was even an error because \textit{``[P3] can't really recall that where it wasn't right.''} Participants who even experienced something odd mentioned that it was difficult for them check whether that was an error, or not. P4 commented on this issue by saying, \textit{``One time [GlAccess] goofed, I think, but I don't know if that [was an error.] I don't know if there were more people in the hallway.''}
Due to the inaccessibility of feedback verification, P10 mentioned being dependent on the system feedback, \textit{``Unless someone was there saying that listening to same thing I was listening to and tell me if it was right or wrong, I had to depend on it.''}
We suspect that these experiences led them to expecting future assistive systems to provide more reliable visual estimates, as P3 commented, \textit{``[Future glasses need to] reassure that it's accurate in telling you like distance [how far] people are.''}

\subsection{Participant Feedback on Teachable Passerby Recognition}
\label{sec:qual_teachable}
Although blind participants did not experience personalizing their smart glasses in the study, they envisioned that such assistive technologies would be able to recognize their family and friends once they provide their photos to the device. For instance, P1 expected this teachable component by saying, \textit{``[Person recognition] means [that] I have to create a database of pictures. And I wouldn't mind doing that if my friends wanted to.''}
Recognizing our lab members during the study seemed to help them see the benefits of recognizing their known people via smart glasses. P6 was excited about using this recognition feature to greet her family and friends by their name even \textit{``when they hadn't spoken to [P6].''}
Also, P10 mentioned training smart glasses to find, \textit{''out of a group, a person that [P10 is] looking for.''} P1 commented on sharing photos with other blind people to recognize each other through such smart glasses, saying \textit{``When I was at the American Council of the Blind convention, there were rooms full of blind people, [and] none of [us] knew who all was in the room. So wouldn't it be cool if we all had each other's pictures [to teach smart glasses about us]?''}

\section{Discussion}

Our exploratory findings from the user study lead to several implications for the design of future assistive systems supporting pedestrian detection, social interactions, and social distancing. We discuss those in this section to stimulate future work on assistive wearable cameras for blind people.

\subsection{Implications regarding Accessing Proxemic Signals using Head-Mounted Cameras}

Section~\ref{sec:quant_visual_info} reveals that some camera frames from blind participants did not capture passersby's face but other body parts (\ie, torso, arms, or legs), especially when they were near the passersby. Although some visual cues such as head pose and facial expressions~\cite{zhao2018face} cannot be extracted from the other body parts, detecting those may help increase accessibility of someone's presence~\cite{yu2018smartpartnet, zhao2018understanding}.
Also, we see that proximity sensors can enhance the detection of a person's presence and distance as those sensors can provide depth information. In particular, such detection can help blind users to practice social distancing, which has posed unique challenges to people with visual impairments~\cite{afb2020flatten}. The sensor data however needs to be processed along with person detection, which often requires RGB data from cameras. Future work that focuses on more robust presence estimation for blind people may consider incorporating proximity sensors with cameras.

Regarding delivery of proxemic signals, it is important to note that blind people may have different spatial perception~\cite{cleaves1979spatial} and could thus interpret such proxemics feedback differently, especially if the feedback was provided with adjectives related to spatial perception, such as \textit{`far'} or \textit{`near'} (Section~\ref{sec:qual_interactions}). Other types of audio feedback (\eg, sonification) or sensory stimulus (\eg, haptic) may overcome the limitations that the speech feedback has.
On the other hand, blind participants' experiences with our testbed may have been affected by its performance and thus could differ if an assistive system had more accurate performance or employed different types of feedback. Future work should consider investigating such a potential correlation.

\subsection{Implications regarding the Form Factor}

\begin{table}[!t]
    \small
    \centering
    \caption{Information about non-handheld cameras designed for helping users with visual impairments access visual surroundings.}
    \begin{tabular}{c c c c}
    \toprule
    Camera Device & FOV & Camera Placement & Design Purpose \\
    \midrule
    Horizon Smart Glasses & ~120-degree horizontal & head-mounted & Aira (agent-support)~\cite{Aira} \\
    OrCam MyEye & \textit{not available} & head-mounted & face recognition \& color identification~\cite{OrCam} \\
    Microsoft HoloLens1 & ~67-degree horizontal & head-mounted & person detection~\cite{stearns2018automated}, stair navigation~\cite{zhao2019designinguist} \\
    Ricoh Theta & ~190-degree horizontal$^a$ & head-mounted & person detection~\cite{stearns2018automated} \\
    ZEDTM2K Stereo & ~90-degree horizontal & suitecase-mounted & person detection~\cite{kayukawa2019bbeep} \\
    Vuzix Blade & ~64-degree horizontal & head-mounted & person detection~\cite{lee2020pedestrian} \\
    RealSense D435 & ~70-degree horizontal & suitcase-mounted & person detection~\cite{kayukawa2020guiding}$^b$ \\
    \bottomrule
    \multicolumn{4}{l}{\footnotesize $^a$It creates a 360-degree panorama by stitching multiple photos.}\\
    \multicolumn{4}{l}{\footnotesize $^b$Two cameras were placed side by side to obtain ~135-degree horizontal POV.}
    \end{tabular}
    \label{tab:cameras_spec}
\end{table}


Several factors, such as blind people's tendency of veering~\cite{guth1994veering} and behavior of scanning the environment when using smart glasses (Section~\ref{sec:quant_visual_info}), can change camera input on wearable cameras and thus may affect the performance of detecting proxemic signals.
From our study, we observed that the head movement of blind people often led to excluding a passerby from camera frames of a head-worn camera.
Although we did not observe blind participants' veering tendency in our study, the veering, which can vary by blind people~\cite{guth1994veering}, could change their camera aiming and consequently affect what is being captured.
Also, as discussed in earlier work~\cite{cleaves1979spatial}, blind people's onset age affects the development of their spatial perception. Observing that our blind participants had different onset ages, we suspect that some differences in their spatial perception may explain their different scanning behaviors with smart glasses. Future work should investigate these topics further to enrich our community's knowledge on these matters.

Moreover, using a wide-angle camera may help assistive systems to capture more visual information (\eg, a passerby coming from the side of a user) and tolerate some variations in blind people's camera aiming with wearable cameras.
The smart glasses in our user study, having only a front view with the limited horizontal range (64-degree angle), are unable to capture and detect people coming from the side or back of a blind user.
As described in Table~\ref{tab:cameras_spec}, some prior work has employed a wide-angle camera~\cite{stearns2018automated} or two cameras side by side~\cite{kayukawa2020guiding} to get more visual data, especially for person detection.
However, accessibility researchers and designers need to investigate which information requires more attention and how it should be delivered to blind users. Otherwise, blind users may be overwhelmed and often distracted by too much feedback (Section~\ref{sec:qual_interactions}).

In addition, aiming with a wearable camera can vary by where the camera is positioned (\eg, a chest-mounted camera) as earlier work on wearable cameras for sighted people indicates that the position of a camera can lead to different visual characteristics of images~\cite{wolf2015effects}. In the context of person detection, prior work investigated assistive cameras for blind people in two different positions (\ie, either on a suitcase~\cite{kayukawa2019bbeep, kayukawa2020guiding} or on a user's head~\cite{stearns2018automated, lee2020pedestrian}). However, little work explores how camera positions can change the camera input while also affecting blind users' mental mapping and computer vision models' performances. We believe that it is important to study how blind people interact with cameras placed in different positions to explore design choices for assistive wearable cameras.

\subsection{Implications regarding the Interactions with the System}

Data-driven systems (\eg, machine learning models) inherently possess the possibility of generating errors due to several factors, such as data shifts~\cite{quinonero2009dataset} and model fragility~\cite{goodfellow2014explaining}.  Since the system that study participants experienced merely served as a testbed, it was naturally error prone (Section~\ref{sec:quant_ret_errors}), which is largely true for any AI system, and consequently affected blind participants' experiences. (Section~\ref{sec:qual_estimates_errors}). In particular, blind participants pointed out inaccessibility of error detection in visual estimates. To help blind users interpret feedback from such data-driven assistive systems effectively, it is imperative to tackle this issue (\ie, \textit{How to make such errors accessible to blind users?}). One simple solution may be to provide the confidence score of an estimation, but further research is needed to understand how blind people interpret this additional information (\ie, confidence score) and to design effective methods of delivering such information.

Furthermore, different feedback modalities are worth exploring. In this work, we provided visual estimates in speech, but blind participants sometimes found it too verbose and imagined that it would be difficult for them to pay attention to speech feedback in certain situations, \eg when they are in noisy areas or talking to someone.
To prevent their auditory sensory system from being overloaded, other sensory stimuli such as haptic can be employed to provide feedback~\cite{azenkot2011smartphone,sanchez2011audio}.
It is however important to understand what information each feedback modality can deliver to blind users and how effectively they would perceive the information.


\subsection{Implications regarding Teachable Interfaces}

Blind participants envisioned personalizing their smart glasses to recognize their family members and friends (Section~\ref{sec:qual_teachable}). To provide this experience for users, it would be worth exploring teachable interfaces, where users provide their own data to teach their systems~\cite{kacorri2017teachable}. Our community has studied data-driven approaches in several applications (\eg, face recognition~\cite{zhao2018face}, object recognition~\cite{kacorri2017people, lee2019hands, ahmetovic2020recog}, and sound recognition~\cite{jain2020soundwatch}) to investigate the potential and implications of using machine learning in assistive technologies. However, little work explores how blind users interact with teachable interfaces, especially for person recognition. Our future work will focus on this to learn opportunities and challenges of employing teachable interfaces in assistive technologies.

Moreover, in Section~\ref{sec:qual_interactions}, blind participants suggested enabling future smart glasses to detect specific moments (\eg, sitting in a table or talking with someone) to control feedback delivery. We believe that egocentric activity recognition~\cite{yan2015egocentric, ma2016going, min2021integrating}, a widely-studied problem in the computer vision community, can help realize this experience. However, further investigation is required for adopting such models in assistive technologies, especially for blind users. For instance, one recent work proposed an egocentric activity recognition model assuming that input data inherently contains the user's gaze motion~\cite{min2021integrating}.  However, that model may not work on data collected by blind people since the assumption, (\ie, eye gaze), only applies to a certain population (\ie, sighted people).

\subsection{Implications regarding Hardware, Privacy Concerns, and Social Acceptance}
There are still challenges related to hardware, privacy, and public acceptance, which keep wearable cameras from being employed for assistive technologies. For example, Aira ended their support for Horizon glasses in March 2020, due to hardware limitations~\cite{Aira2020stop}.
Also, prior work investigated blind users' privacy concerns over capturing their environments without being given access to what is being captured~\cite{ahmed2016addressing, akter2020privacy}, and social acceptance of assistive wearable cameras as the always-on cameras can capture other people without notification~\cite{profita2016effect, ahmed2018up, lee2020pedestrian}.

Beyond the high cost of real-time support from sighted people (\eg, Aira), as of now, there is no effective solution that can estimate the distance between blind users and other people in real time.
Computer vision has the potential to provide this proxemic signal in real time on a users' device. No recordings~\cite{lee2020pedestrian} and privacy-preserving face recognition~\cite{erkin2009privacy, ren2018learning} may help address some privacy issues, but more concerns may arise if assistive wearable cameras need to store visual data (even on the device itself) to recognize the users' acquaintances. Future work should consider investigating associated privacy concerns and societal issues, as these factors can affect design choices for assistive wearable cameras.


\section{Conclusion}
In this paper, we explored the potential and implications of using a head-worn camera to help blind people access proxemics, such as the presence, distance, position, and head pose, of pedestrians.
We built GlAccess that focused on detecting the head of a passerby to estimate proxemic signals. We collected and annotated camera frames and usage logs from ten blind participants, who walked in a corridor while wearing the smart glasses for pedestrian detection and interacting with a total of 80 passersby.
Our analysis results show that their smart glasses tended to capture passersby's head when they were more than 6 feet (2 meters) away from passersby. However, the rate of including the head dropped quickly as blind participants were getting closer to passersby, where we observed variations in their scanning behaviors (\ie, head movements).
Blind participants shared that using smart glasses for pedestrian detection was easy since they did not have to manipulate camera aiming. Their qualitative feedback also led to several design implications for future assistive cameras to consider. In particular, we found that errors in visual estimates need to be accessible to blind users to better interpret such outputs for their autonomy.
Our future work will explore this accessibility issue in several use cases where blind people can benefit from wearable cameras.

\begin{acks}
We thank Tzu-Chia Yeh for her help in our work presentation and the anonymous reviewers for their constructive feedback on an earlier draft of our work. This work is supported by NIDILRR (\#90REGE0008) and Shimizu Corporation.
\end{acks}

\bibliographystyle{ACM-Reference-Format}
\bibliography{main}

\end{document}